\documentclass[12pt]{article}
\usepackage{epsf,epsfig,amssymb,amsmath,graphicx,float}
\usepackage{color}

\newcommand{\lsim}{\raisebox{-0.13cm}{~\shortstack{$<$ \\[-0.07cm] $\sim$}}~}

\begin{document}
\renewcommand{\thefootnote}{\fnsymbol{footnote}}

\begin{titlepage}

\begin{center}

\vspace{1cm}

{\Large {\bf Relic Abundance of Asymmetric Dark Matter in Quintessence }}

\vspace{1cm}

{\bf Hoernisa Iminniyaz}$^a$\footnote{wrns@xju.edu.cn},
{\bf Xuelei Chen}$^b$\footnote{xuelei@cosmology.bao.ac.cn}\\
\vskip 0.15in
{\it
$^a${School of Physics Science and Technology, Xinjiang University, \\
Urumqi 830046, China} \\
$^b${National Astronomical Observatories, Chinese Academy of
    Sciences,\\ Beijing 100012, China}\\
}

\abstract{We investigate the relic abundance of asymmetric Dark Matter
  particles in quintessence model with a kination phase. The analytic
  calculation of the asymmetric Dark Matter in the standard cosmological
  scenario is extended to the nonstandard cosmological scenario where we
  specifically discuss the quintessence model with a kination phase. We found
  that the enhancement of Hubble rate changes the relic density of
  particles and anti--particles. We use the present day Dark Matter abundance
  to constrain the Hubble rate in quintessence model with a kination phase for
  asymmetric Dark Matter.}
\end{center}
\end{titlepage}
\setcounter{footnote}{0}

\section{Introduction}

The cosmological and astrophysical observations showed
that the universe contains large amount of Dark Matter. The Dark
Matter relic density is determined by Cosmic Microwave Background
(CMB) Anisotropy observations with the Wilkinson Microwave
Anisotropy Probe (WMAP) as \cite{wmap},
\begin{eqnarray} \label{wmap}
  \Omega_{\rm DM} h^2 = 0.1109 \pm 0.0056\, ,
\end{eqnarray}
where $\Omega_{\rm DM}$ is the Dark Matter (DM) density in unit of the
critical density, and $h = 0.710 \pm 0.025$ is the Hubble constant in
units of 100 km sec$^{-1}$ Mpc$^{-1}$.

The nature of the Dark Matter is still a challenging questions for
scientists though we have a precise measurement of Dark Matter amount.
So far, many Dark Matter candidate particles have been proposed
beyond the Standard Model (SM). Among them neutral, long--lived or
stable weakly interacting massive particles (WIMP) are considered
excellent candidates for Dark Matter. Neutralino is one of the most
promising candidate for Dark Matter which appears in
supersymmetric standard model, neutralino is stabilized due to the
conserved $R$--parity \cite{review}. Neutralino is Majorana particle
for which its particle and anti-particle are the same. However this
is only one possibility. Most of the known elementary particles in
the universe indeed are not Majorana particles, the particles and
anti--particles are distinct if we consider fermionic particles. 
There is one option that
assumes the Dark Matter particles can be Dirac particles. The average
density of baryons and Dark Matter is comparable. This motivates to
consider Dark Matter can be asymmetric particles
\cite{adm-models,frandsen}. \cite{Iminniyaz:2011yp, GSV} investigated the
relic abundance of asymmetric Dark Matter particles in the standard
cosmological scenario in which particles were in thermal equilibrium
in the early universe and decoupled when they were
non--relativistic.

On the other hand, the Dark Matter relic density is changed by the modified
expansion rate of the universe. The reason for that might be the additional
contributions to the total energy density from quintessence model
\cite{Salati:2002md},
anisotropic expansion \cite{Kamionkowski}, a modification of general
relativity \cite{Kamionkowski,Catena} and etc. In
paper by Salati \cite{Salati:2002md}, it was shown that the relic density of
Dark Matter is increased when the expansion rate of the universe is changed in
the quintessence model. There is no
discussion about the asymmetric Dark Matter relic density in nonstandard
cosmological scenarios including quintessence model until now. It deserves
to investigate the relic density of asymmetric Dark Matter in the nonstandard
cosmological scenarios and find to what extent the asymmetric Dark Matter relic
density is affected by the modification of the Hubble rate.

In this paper, we extend the discussion about the relic density of asymmetric
WIMP Dark Matter in the standard cosmological scenario to the nonstandard
cosmological scenario, specifically we examine the asymmetric WIMP Dark Matter
relic density in quintessence model with a kination phase. We assume that the
Dark Matter asymmetry is created before Dark Matter annihilation reactions
freeze--out. In the beginning we assume there are more particles than the
anti--particles. We find
that the enhanced Hubble rate in quintessence model changes the relic
density of both particles and anti--particles. We closely follow the analytic
solution of asymmetric Dark Matter in standard cosmological scenario and
derived the analytic solution of the relic density of asymmetric Dark Matter in
quintessence model.

This paper is arranged as follows. In section 2, we review the relic density
of asymmetric Dark Matter in the standard cosmological scenario. In
section 3, the relic density of asymmetric Dark Matter in quintessence model
with a kination phase is discussed. In section 4, we constrained the
expansion rate in quintessence model with a kination phase using the observed
Dark Matter abundance. The last section is devoted to the conclusions and
discussions.

\section{Relic Abundance of Asymmetric Dark Matter in the Standard
  Cosmological Scenario}

In this section, we review the relic abundance of asymmetric Dark Matter
in the standard cosmological scenario which assumes particles were in thermal
equilibrium in the early universe and decoupled when they were
non--relativistic \cite{Iminniyaz:2011yp}. $\chi$ is denoted as a Dark Matter
particle that is {\em not} self--conjugate, i.e. the anti--particle
$\bar\chi \neq \chi$. Time evolutions of the number densities
$n_{\chi}$, $n_{\bar\chi}$ in the expanding universe are described by the
Boltzmann equations. Solving Boltzmann equations, we obtain the relic
densities of $\chi$ and $\bar\chi$
particles. It is assumed that only $\chi \bar \chi$ pairs can
annihilate into Standard Model (SM) particles, while $\chi\chi$ and
$\bar \chi \bar\chi$ pairs can not, the Boltzmann equations are:
\begin{eqnarray} \label{eq:boltzmann_n}
\frac{{\rm d}n_{\chi}}{{\rm d}t} + 3 H n_{\chi} &=&  - \langle \sigma v\rangle
  (n_{\chi} n_{\bar\chi} - n_{\chi,{\rm eq}} n_{\bar\chi,{\rm eq}})\,;
  \nonumber \\
\frac{{\rm d}n_{\bar\chi}}{{\rm d}t} + 3 H n_{\bar\chi} &=&
   - \langle \sigma v\rangle (n_{\chi} n_{\bar\chi} - n_{\chi,{\rm
       eq}} n_{\bar\chi,{\rm eq}})\,,
\end{eqnarray}
where $\langle \sigma v \rangle$ is the thermally averaged
annihilation cross section multiplied with the relative velocity of the two
annihilating $\chi$, $\bar{\chi}$ particles. $H = \dot{R}/R$ is the expansion
rate of the universe, where $R$ is the scale factor of the universe.
$n_{\chi,{\rm eq}}$, $n_{\bar{\chi},{\rm eq}}$ are the equilibrium number
densities of $\chi$ and $\bar{\chi}$, here it is assumed the Dark Matter
particles were non--relativistic at decoupling. Then the equilibrium number
densities $n_{\chi,{\rm eq}}$ and $n_{\bar\chi,{\rm eq}}$ are
\begin{eqnarray} \label{n_eq}
  n_{\chi,{\rm eq}} &=& g_\chi ~{\left( \frac{m_\chi T}{2 \pi} \right)}^{3/2}
  {\rm e}^{(-m_\chi + \mu_\chi)/T}\,, \nonumber \\
  n_{\bar\chi,{\rm eq}} &=&  g_\chi ~{\left( \frac{m_\chi T}{2 \pi}
    \right)}^{3/2} {\rm e}^{(-m_\chi - \mu_{\bar\chi})/T}\,,
\end{eqnarray}
where $m_\chi$ is the mass of the Dark Matter particle $\chi$ and
$\bar{\chi}$, and $g_{\chi}$ is the number of the internal
degrees of freedom of the $\chi$ and $\bar{\chi}$ separately.
$\mu_\chi$, $\mu_{\bar\chi}$ are the chemical potential of the particles and
anti--particles, $\mu_{\bar\chi} = -\mu_\chi$ in equilibrium.

At high temperature $\chi$ and $\bar\chi$
particles are in thermal equilibrium in the early universe. When
$T < m_\chi$, for $m_\chi > |\mu_\chi|$, the number densities
$n_{\chi,{\rm eq}}$ and $n_{\bar\chi,{\rm eq}}$ decrease exponentially.
Finally the interaction rates $\Gamma = n_{\chi}
\langle \sigma v \rangle$ and $\bar{\Gamma} = n_{\bar\chi} \langle \sigma v
\rangle$ drop below $H$. $\chi$ and $\bar\chi$ particles
are then no longer kept in chemical equilibrium, and their co--moving number
densities are fixed. The temperature at which the WIMPs drop
out of chemical equilibrium is called the freeze--out temperature.

For convenient, the Boltzmann equations (\ref{eq:boltzmann_n})
can be rewritten in terms of the dimensionless quantities $Y_\chi = n_\chi/s$,
$Y_{\bar\chi} = n_{\bar\chi}/s$, and $x = m_\chi/T$. The entropy density is
given by $ s= (2 \pi^2/45) g_{*s} T^3 $, where
\begin{eqnarray}
  g_{*s} = \sum_{i = {\rm bosons}} g_i \left( \frac{T_i}{T} \right)^3
  + \frac{7}{8} \sum_{i = {\rm fermions}} g_i
  \left( \frac{T_i}{T} \right)^3\, .
\end{eqnarray}
Here $g_i$ is equivalent to $g_{\chi}$, $T_i$ is the temperature of
species $i$. During the radiation--dominated epoch, the expansion rate $H$
is given by
\begin{equation} \label{H}
H = \frac{\pi T^2}{M_{\rm Pl}} \sqrt{\frac{g_*}{90}}\,,
\end{equation}
with $M_{\rm Pl} = 2.4 \times 10^{18}$ GeV being the reduced Planck mass
and
\begin{eqnarray}
  g_{*} = \sum_{i = {\rm bosons}} g_i \left( \frac{T_i}{T} \right)^4
  + \frac{7}{8} \sum_{i = {\rm fermions}} g_i
  \left( \frac{T_i}{T} \right)^4\, .
\end{eqnarray}  
It is assumed that the universe expands adiabatically during this period,
then the Boltzmann equations (\ref{eq:boltzmann_n}) become
\begin{equation} \label{eq:boltzmann_Y}
\frac{d Y_{\chi}}{dx} = - \frac{\lambda \langle \sigma v \rangle}{x^2}~
(Y_{\chi}~ Y_{\bar\chi} - Y_{\chi, {\rm eq}}~Y_{\bar\chi, {\rm eq}}   )\,;
\end{equation}
\begin{equation} \label{eq:boltzmann_Ybar}
\frac{d Y_{\bar{\chi}}}{dx}
= - \frac{\lambda \langle \sigma v \rangle}{x^2}~
 (Y_{\chi}~Y_{\bar\chi} - Y_{\chi, {\rm eq}}~Y_{\bar\chi, {\rm eq}} )\,,
\end{equation}
where
\begin{equation} \label{lambda}
\lambda = 1.32\,m_{\chi} M_{\rm Pl}\, \sqrt{g_*}\,.
\end{equation}
Here we assume $g_* \simeq g_{*s}$ and ${\rm d}{g_*}/{\rm d}x
\simeq 0$.

Subtracting Eq.(\ref{eq:boltzmann_Y}) from Eq.(\ref{eq:boltzmann_Ybar}),
\begin{equation} \label{eq:YYbar}
\frac{d Y_{\chi}}{dx} - \frac{d Y_{\bar{\chi}}}{dx} = 0\,.
\end{equation}
This indicate
\begin{equation}  \label{eq:c}
 Y_{\chi} - Y_{\bar\chi} = C\,,
\end{equation}
where $C$ is a constant, the difference of the co--moving densities
of the particles and anti--particles is conserved. Expressing
Eqs.(\ref{eq:boltzmann_Y}) and (\ref{eq:boltzmann_Ybar}) using
Eq.(\ref{eq:c}), the Boltzmann equations become
\begin{equation} \label{eq:Yc}
\frac{d Y_{\chi}}{dx} = - \frac{\lambda \langle \sigma v \rangle}{x^2}~
      (Y_{\chi}^2 - C Y_{\chi} - P     )\, ;
\end{equation}
\begin{equation} \label{eq:Ycbar}
\frac{d Y_{\bar{\chi}}}{dx} = - \frac{\lambda \langle \sigma v \rangle}{x^2}~
 (Y_{\bar\chi}^2 + C Y_{\bar\chi}  - P)\,,
\end{equation}
where
\begin{equation} \label{P}
P = Y_{\chi,{\rm eq}} Y_{\bar\chi,{\rm eq}} = (0.145
g_{\chi}/g_*)^2\,x^3\,{\rm e}^{-2x}\,.
\end{equation}

Usually the WIMP annihilation cross section can be expanded in
the relative velocity $v$ between the annihilating WIMPs. Its thermal average
is given by
\begin{equation} \label{eq:cross}
   \langle \sigma v \rangle = a + 6\,b x^{-1} + {\cal O}(x^{-2})\, .
\end{equation}
Here $a$ is $s$--wave contribution to $\sigma v$ while $b = 0$, and
$b$ is $p$--wave contribution to $\sigma v$ while $a = 0$.

In the standard cosmological scenario, the following approximate formulae
are obtained for the relic abundances of particles and anti--particles:

\begin{equation} \label{eq:Yq_cross}
       Y_{\chi}(x \rightarrow \infty) =   \frac{C}
       { 1 - \exp \left[- 1.32\, C\, m_{\chi} M_{\rm Pl}\,
        \sqrt{g_*}  \, (a x_F^{-1} + 3 b x_F^{-2})  \right]}\,,
\end{equation}
\begin{equation} \label{eq:barYq_cross}
Y_{\bar\chi}(x \rightarrow \infty) =  \frac{C}
 { \exp \left[ 1.32\, C \, m_{\chi} M_{\rm Pl}\,
 \sqrt{g_*} \, (a~ \bar{x}_F^{-1} + 3 b~ \bar{x}_F^{-2})   \right] -1}\,,
\end{equation}
where we have used Eq.(\ref{lambda}). $x_F$ and $\bar{x}_F$ are the inverse
scaled freeze--out temperatures of $\chi$ and $\bar{\chi}$.
Eqs.(\ref{eq:Yq_cross}) and (\ref{eq:barYq_cross}) are only consistent with
the constraint (\ref{eq:c}) if $x_F = \bar x_F$. For convenience, the final
abundance is expressed as
\begin{equation}
\Omega_\chi h^2 =\frac{m_\chi s_0 Y_{\chi}(x \to \infty) h^2}{\rho_{\rm
  crit}}\,,
\end{equation}
where $s_0 = 2.9 \times 10^3~{\rm cm}^{-3}$ is the present
entropy density, and $\rho_{\rm crit} = 3 M_{\rm Pl}^2 H_0^2$ is the present
critical density. The corresponding prediction for the present relic density
for Dark Matter is given by
\begin{equation} \label{omega}
\Omega_{\rm DM} h^2 = 2.76 \times 10^8~ m_\chi \left[ Y_{\chi}~(x
  \rightarrow \infty) + Y_{\bar\chi}~(x \rightarrow \infty) \right]
\mbox{GeV}^{-1}\, .
\end{equation}
We defer further discussions of this expression to Sec.3 where the asymmetric
Dark Matter relic density in the quintessence model with a
kination phase is analyzed.

\section{Relic Abundance of WIMPs in Quintessence}

Before going to the discussion of the relic density in quintessence
model with a kination phase, let us briefly review the model.
A kination is a period in which the kinetic energy of a scalar
field $\rho_{\phi} \simeq \dot{\phi}^2/2$
dominates over the potential energy density $V(\phi)$ and the radiation
energy density $\rho_{\it rad}$. A kination phase
is appeared in quintessence models based on tracking solutions for
the scalar field \cite{Salati:2002md}. The overall energy density decreases
as $\rho_{\it tot} \simeq \dot{\phi}^2/2 \sim R^{-6}$. $T \sim R^{-1}$, thus
$H^2 \sim \rho_{tot} \sim T^6 $. It means $H$ decreases faster as $T$
decreases or $H$ decreases faster as the scale factor $R$ increases.
The ratio of the expansion rate $H$ during kination period and the expansion
rate of the standard case $H_{\rm std}$ is given by:
\begin{equation}\label{Hratio}
      \frac{H^2}{H^2_{\rm std}} = 1 + \frac{\rho_{\phi}}{\rho_r}\, ,
\end{equation}
where the ratio of the scalar energy density to the radiation energy density
$\rho_{\phi}/\rho_r$ is \cite{Salati:2002md}
\begin{equation}\label{rhoratio}
      \frac{\rho_{\phi}}{\rho_r} = \eta
      \left[ \frac{g_{*s}(T)}{g_{*s}(T_0)} \right]^2~ \frac{g_*(T_0)}{g_*(T)}~
       \left( \frac{T}{T_0} \right)^2 \simeq \eta
       \left( \frac{T}{T_0}  \right)^2\, ,
\end{equation}
where $\eta = \rho_{\phi}(T_0)/\rho_r(T_0)$. Here $T_0$ is some reference
temperature which is close to the freeze--out temperature. The approximation
of the above equation only holds in a range of temperatures where
$g_{*s}$ and $g_*$ do not change sizably with respect to their value
at $T_0$.
We can rewrite Eq.(\ref{Hratio}) as
\begin{eqnarray}
      H & = & A(T) H_{\rm std},
\end{eqnarray}
where the enhancement function $A(T)$ is
\begin{equation}\label{AT}
      A(T) = \sqrt{1 + \eta~\left( \frac{T}{T_0} \right)^2}\,.
\end{equation}
In order not to spoil the successful prediction of Big Bang
Nucleosynthesis BBN, $A(T)$
must return to 1 at the low temperature around 1 MeV.
With the modified expansion rate, the Boltzmann Eqs.(\ref{eq:Yc}),
( \ref{eq:Ycbar} ) become
\begin{equation} \label{eq:Ycq}
\frac{d Y_{\chi}}{dx} = - \frac{\lambda \langle \sigma v \rangle}{x^2 A(x)}~
      (Y_{\chi}^2 - C Y_{\chi} - P     )\, ;
\end{equation}
\begin{equation} \label{eq:Ycbarq}
\frac{d Y_{\bar{\chi}}}{dx} = -\frac{\lambda \langle \sigma v \rangle}{x^2 A(x)}~
 (Y_{\bar\chi}^2 + C Y_{\bar\chi}  - P)\,.
\end{equation}
\begin{figure}[h!]
  \begin{center}
    \hspace*{-0.5cm} \includegraphics*[width=9cm]{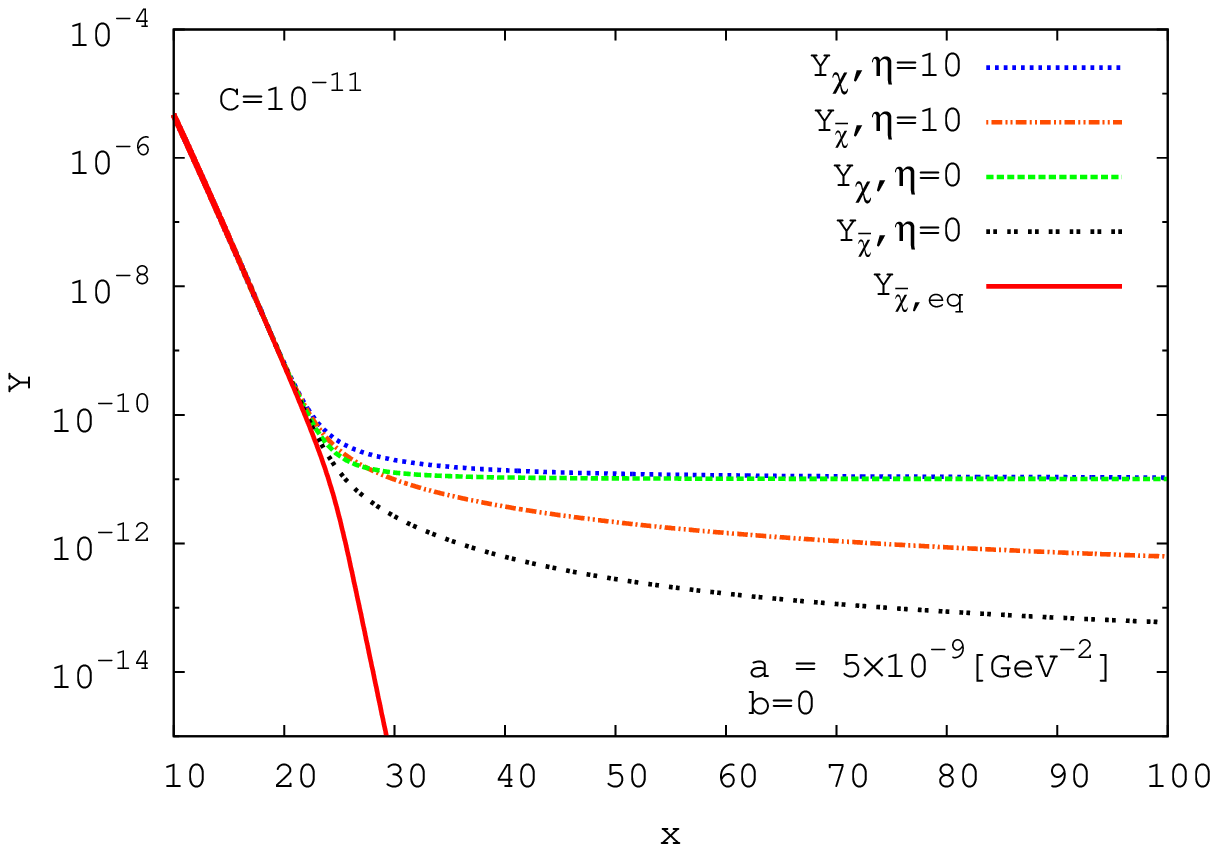}
    \put(-115,-12){(a)}
    \hspace*{-0.5cm} \includegraphics*[width=9cm]{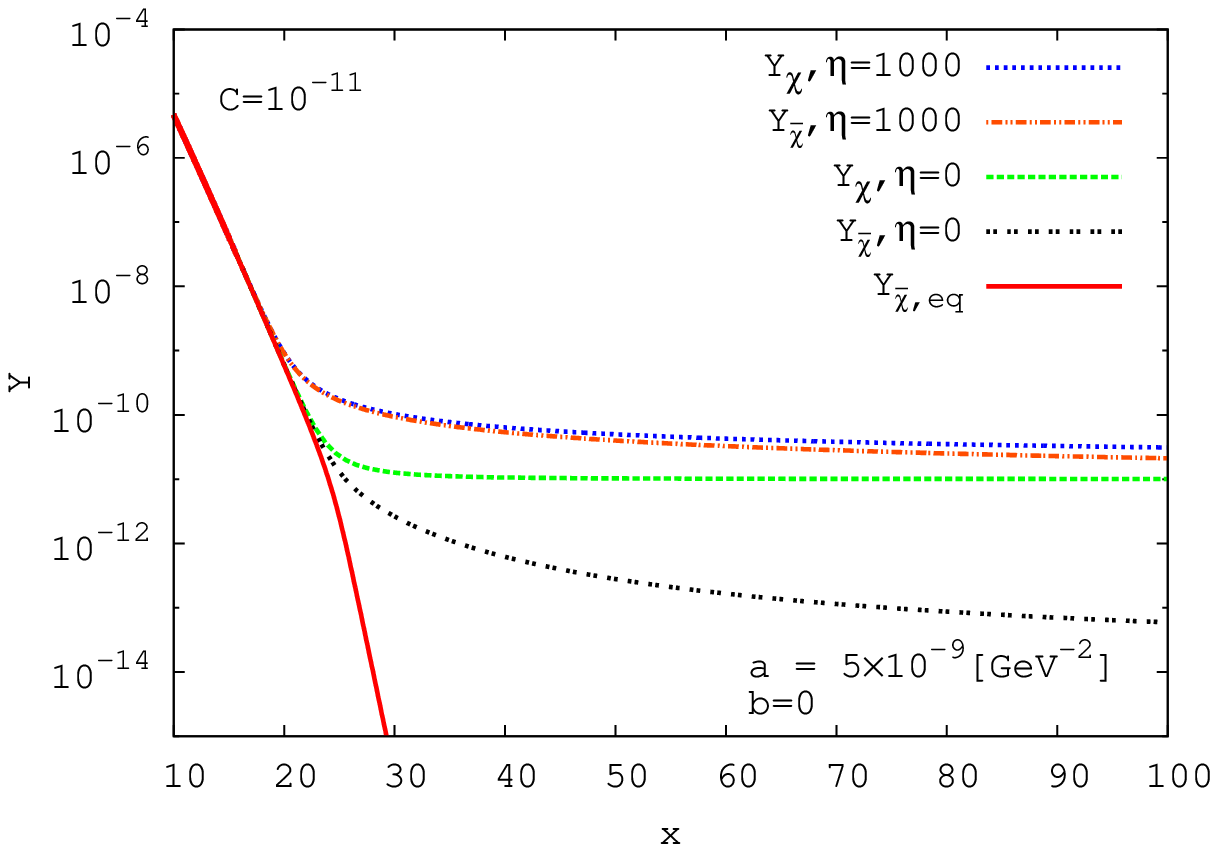}
    \put(-115,-12){(b)}
     \vspace{0.5cm}
    \hspace*{-0.5cm} \includegraphics*[width=9cm]{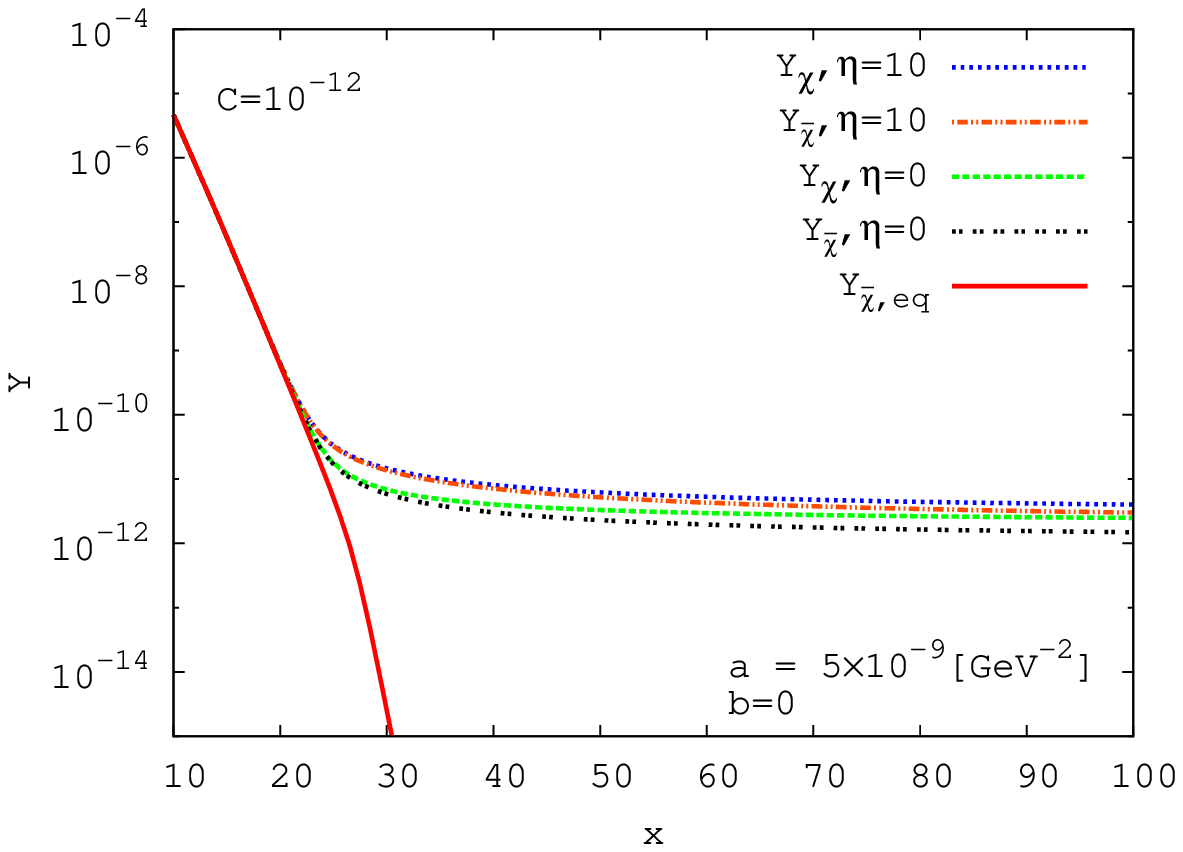}
    \put(-115,-12){(c)}
    \hspace*{-0.5cm} \includegraphics*[width=9cm]{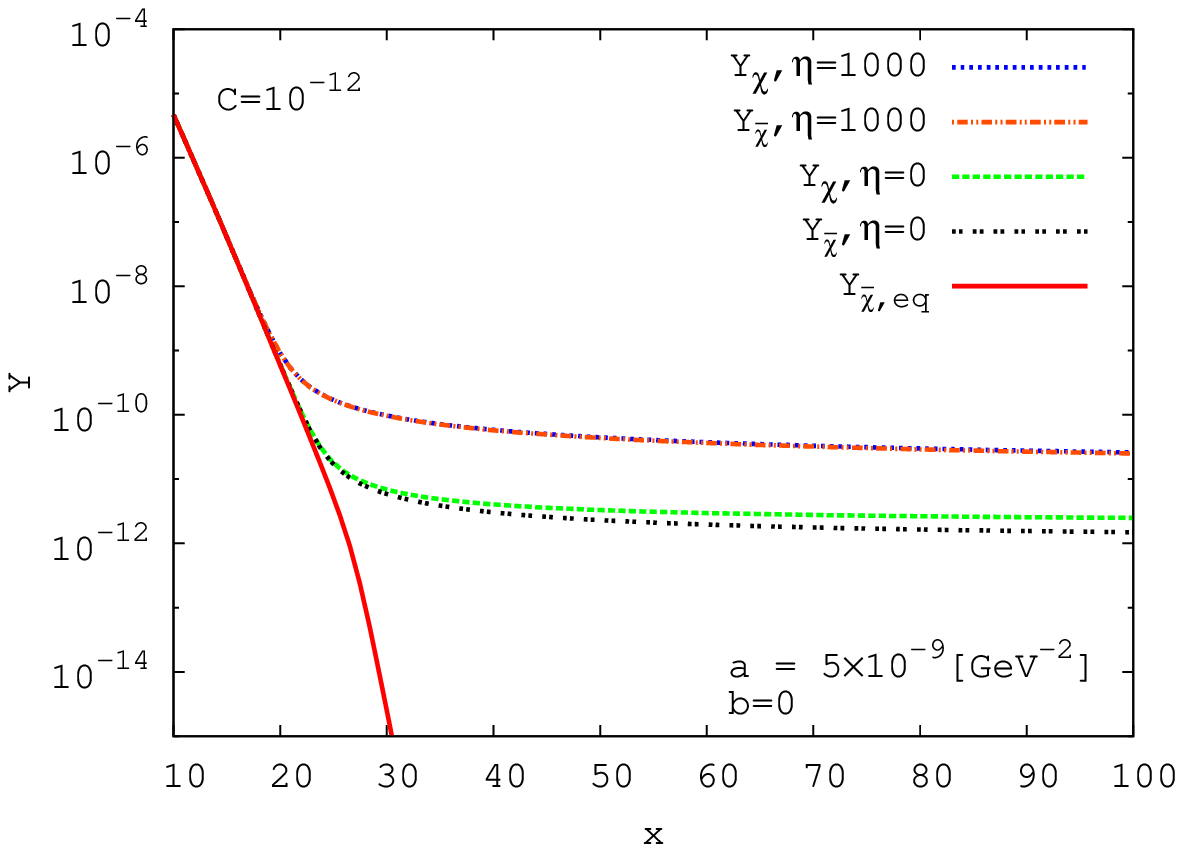}
    \put(-115,-12){(d)}
    \caption{\label{fig:a} \footnotesize The relic abundances $Y_{\chi}$ and
     $Y_{\bar\chi}$ for particle $\chi$ and anti--particle $\bar\chi$ as a
     function of the inverse--scaled temperature for $\eta = 10$ (a), (c) and
     $\eta = 1000$ (b), (d). Here we take $a = 5.0 \times 10^{-9}$ GeV$^{-2}$
     $ = 5.9 \times 10^{-26} $ cm$^3$ s$^{-1}$,
     $b=0$, $m = 100$ GeV,  $x_0 = 20$, $C = 10^{-11}$ for (a) and (b),
     $C = 10^{-12}$ for (c) and (d) }
  \end{center}
\end{figure}
Fig.\ref{fig:a} is obtained by the numerical solutions of
Eqs.(\ref{eq:Ycq}), (\ref{eq:Ycbarq}) for the different values of the
enhancement factor $\eta$ with different asymmetry factor $C$.
The solid (red) line is the equilibrium
value of the anti--particle abundance. The double dotted (black)
line is for the anti--particle abundance and the dashed (green) line
is for particle abundance in the standard scenario ($\eta =0$). The
dot--dashed (red) line is for the abundance of anti--particle and
dotted (blue) line is for abundance of particle for $\eta = 10$
left (a), (c) frames and $\eta = 1000$ right (b), (d) frames. In
\cite{Schelke:2006eg, Donato:2006af, Catena:2009tm}, the authors found
the constraints for $\eta$ as $\eta < 10^6$ for the WIMPs mass
$m_{\chi} < 1$ TeV. We discuss the asymmetric WIMPs mass $m_{\chi} < 1$ TeV.
Therefore, we take $\eta = 10$ and $\eta = 1000$ for examples. 
Here we take $ C = 10^{-11} $, for (a), (b) and $C = 10^{-12}$ for
$(c), (d)$, $m_\chi = 100$ GeV, $a = 5.0 \times 10^{-9}$ GeV$^{-2}$ $ = 5.9 \times 10^{-26} $
cm$^3$ s$^{-1}$, $b = 0$, $g_{\chi} = 2$
and $g_* = 90$, $x_0 = 20$, where $x_0 = m/T_0$. 

The relic abundances $Y_{\chi}$ and $Y_{\bar\chi}$ for particle $\chi$
and anti--particle $\bar\chi$ are increased in Fig.\ref{fig:a} (a), (c) frames for $\eta = 10$
and (b), (d) frames for $\eta = 1000$. The expansion rate is enhanced in quintessence
model with a kination phase. The particles and anti--particles freeze--out
earlier than the standard case. This leads to the increases of particle and
anti--particle abundances. As we noted in introduction, we assume
there are more particles than the anti--particles.
For smaller $\eta$ ($\eta = 10$, (a), (c) left frames), the relic
abundances for particles and anti--particles are not affected too much.
When $\eta$ is large ($\eta = 1000$, (b), (d) right frames), the relic
abundances for particles and anti--particles are increased significantly.
In frames (a), (b), the relic abundance of anti--particle
seems to be increased more sizably than the particle relic abundance for the
asymmetry factor $C = 10^{-11}$. In fact
there are same increases for particle and anti--particle relic abundances.
The reason is the following: let's closely look at the left (a) frame of
Fig.\ref{fig:a}. The anti--particle relic abundance
$Y_{\bar\chi}$ is $8.0 \times 10^{-14}$ for $\eta =0$. It is
increased to $8.0 \times 10^{-13}$ for $\eta = 10$. It is one order increase.
According to Eq.(\ref{eq:c}), $Y_{\chi} = Y_{\bar \chi} + C$, here $C =
10^{-11}$. Thus the particle $\chi$ abundance is increased almost
$10\%$ for $\eta = 10$. This is only small deviation. On the other hand, when the asymmetry factor
$C$ is small ($C = 10^{-12}$), the increases for particles and anti--particles are comparable
which are shown in frames (c) and (d) in Fig.\ref{fig:a}.

Following \cite{Iminniyaz:2011yp}, we obtain the analytic solution of the
relic density for asymmetric Dark Matter in quintessence model with a kination
phase. First, we solve Eq.(\ref{eq:Ycbarq}) for $\bar{\chi}$ density, then
$\chi$ density can be computed trivially using Eq.(\ref{eq:c}). We introduce
the quantity $\Delta_{\bar\chi} = Y_{\bar\chi} - Y_{\bar\chi,{\rm eq}}$. In
terms of $\Delta_{\bar\chi}$, the Boltzmann equation (\ref{eq:Ycbarq}) can
be rewritten as:
\begin{equation} \label{eq:delta}
\frac{d \Delta_{\bar\chi}}{dx} = - \frac{d Y_{\bar\chi,{\rm eq}}}{dx} -
\frac{\lambda \langle \sigma v \rangle}{x^2 A(x)}~
\left[\Delta_{\bar\chi}(\Delta_{\bar\chi} + 2 Y_{\bar\chi,{\rm eq}})
      + C \Delta_{\bar\chi}   \right]\, .
\end{equation}
The solution of this equation can be considered in two
regimes. At sufficiently high temperature, $Y_{\bar\chi}$ tracks its equilibrium
value $Y_{\bar\chi,{\rm eq}}$ very closely. In that regime $\Delta_{\bar\chi}$ is
small, and $d \Delta_{\bar\chi}/dx$ and $\Delta_{\bar\chi}^2$ are
negligible. The Boltzmann equation (\ref{eq:delta}) then becomes
\begin{equation}\label{eq:delta_simp}
     \frac{d Y_{\bar\chi,{\rm eq}}}{dx}   =  -
      \frac{\lambda \langle \sigma v \rangle}{x^2 A(x)}~
      \left(2 \Delta_{\bar\chi} Y_{\bar\chi,{\rm eq}} +
       C \Delta_{\bar\chi} \right)\,.
\end{equation}
We need an explicit expression for the equilibrium density
$Y_{\bar\chi,{\rm eq}}(x)$ to solve Eq.(\ref{eq:delta_simp}).
The right--hand sides of the Boltzmann equations (\ref{eq:boltzmann_Y}) and
(\ref{eq:boltzmann_Ybar}) vanish in equilibrium by definition. Hence the
right--hand side of Eq.(\ref{eq:Ycbarq}) should vanish as well for
$Y_{\bar\chi} = Y_{\bar\chi,{\rm eq}}$, which implies
\begin{equation} \label{eq:equili}
Y_{\bar\chi,{\rm eq}}^2 + C Y_{\bar\chi,{\rm eq}} - P   = 0\,.
\end{equation}
There are two solutions for the quadratic equation, only one of them yields
a positive $\bar\chi$ equilibrium density:
\begin{equation} \label{eq:bar_eq}
      Y_{\bar\chi,{\rm eq}} = - \frac{C}{2} + \sqrt{\frac{C^2}{4} + P}\,.
\end{equation}
Eq.(\ref{eq:bar_eq}) is inserted into Eq.(\ref{eq:delta_simp}) and
ignoring $x$ relative to $x^2$, we have
\begin{equation} \label{bardelta_solu}
      \Delta_{\bar\chi} \simeq \frac{2 x^2 A(x) P}
      {\lambda \langle \sigma v \rangle\,(C^2 + 4 P)}\,.
 \end{equation}
We will use this solution to determine the freeze--out temperature $\bar{x}_F$
for $\bar{\chi}$.

When the temperature is sufficiently low, i.e. for $x > \bar x_F$,
the production term $\propto Y_{\bar\chi,{\rm eq}}$ can be ignored
in the Boltzmann equation (\ref{eq:delta}), so that
\begin{equation} \label{eq:delta_late}
\frac{d \Delta_{\bar\chi}}{dx} = - \frac{\lambda \langle \sigma v
\rangle}{x^2 A(x)} \left( \Delta_{\bar\chi}^2 + C \Delta_{\bar\chi}    \right)\,.
\end{equation}
Integrating Eq.(\ref{eq:delta_late}) from
$\bar{x}_F$ to $\infty$ and assuming $\Delta_{\chi}(\bar{x}_F) \gg
\Delta_{\chi}(\infty) $, we have
\begin{equation} \label{eq:barY_cross}
Y_{\bar\chi}(x \rightarrow \infty) =  \frac{C}
 { \exp \left[ 1.32\, C \, m_{\chi} M_{\rm Pl}\,
 \sqrt{g_*} \, I(\bar{x}_F)   \right] -1}\,,
\end{equation}
here
\begin{eqnarray}
   I(\bar{x}_F) & =& \int^{\infty}_{\bar{x}_F} \frac{ \langle \sigma v \rangle }
              {x^2~ A(x)}~ dx \\ &=& \frac{a}{\sqrt{\eta}~ x_0} {\rm ln}
               \left( \sqrt{\eta}\, \frac{x_0}{\bar{x}_F} +
              \sqrt{1 + \eta\, \frac{x_0^2}{\bar{x}_F^2} }\,    \right)  +
              \frac{6b}{\eta x^2_0}
              \left( \sqrt{1 + \eta \frac{x_0^2}{\bar{x}^2_F}} - 1  \right)\,.
\end{eqnarray}
Using equation (\ref{eq:c}), we obtain the relic abundance
for $\chi$ particle. The result is
\begin{equation} \label{eq:Y_cross}
       Y_{\chi}(x \rightarrow \infty) =   \frac{C}
       { 1 - \exp \left[-1.32\, C \, m_{\chi} M_{\rm Pl}\,
        \sqrt{g_*}  \, I(x_F)  \right]}\,,
\end{equation}
where $I(x_F)$ is given by
\begin{eqnarray}
    I(x_F) & =& \int^{\infty}_{x_F} \frac{ \langle \sigma v \rangle }
              {x^2~ A(x)}~ dx \\ &=& \frac{a}{\sqrt{\eta}~ x_0} {\rm ln}
              \left( \sqrt{\eta}\, \frac{x_0}{x_F} +
              \sqrt{1 + \eta\, \frac{x_0^2}{x_F^2} }\, \right)  +
              \frac{6b}{\eta x^2_0}
              \left( \sqrt{1 + \eta \frac{x_0^2}{x^2_F}} - 1 \right)\,.
\end{eqnarray}
Here Eqs.(\ref{eq:barY_cross}) and (\ref{eq:Y_cross}) are
only consistent with the constraint (\ref{eq:c}) if $x_F = \bar
x_F$. The prediction for the present relic density for Dark Matter is then
given by
\begin{eqnarray} \label{omega}
 \Omega_{\rm DM}  h^2  =
    \frac{2.76 \times 10^8~ m_\chi ~C}
 { \exp \left[ 1.32 \, C \, m_{\chi} M_{\rm Pl}\,
 \sqrt{g_*} \, I(\bar{x}_F)   \right] -1}\,
      + \frac{2.76 \times 10^8~ m_\chi ~C}
       { 1 - \exp \left[-1.32\, C \, m_{\chi} M_{\rm Pl}\,
        \sqrt{g_*}  \, I(x_F)  \right]}\,.
\end{eqnarray}
When $A(x) = 1$, the standard result for asymmetric Dark Matter is recovered.
The freeze--out temperature for $\bar{\chi}$ is fixed
by assuming that freeze--out occurs when the deviation
$\Delta_{\bar\chi} $ is of the same order of the equilibrium value of
$Y_{\bar\chi}$:
\begin{equation} \label{xf1}
\xi Y_{\bar\chi,{\rm eq}}( \bar{x}_{F_0}) = \Delta_{\bar\chi}( \bar{x}_{F_0})\,,
\end{equation}
where $\xi$ is a numerical constant of order unity,
$\xi = \sqrt{2} -1$ \cite{standard-cos}. $\bar{x}_{F_0}$ is the
freeze--out temperature which is calculated from Eq.(\ref{xf1}) using the
standard approximation. We found the standard treatment under--predicts the
$\bar{\chi}$ relic density. Therefore, the correction is made for the
freeze--out temperature as following:
\begin{equation} \label{eq:xF}
\bar{x}_F = \bar{x}_{F_0}\,  \left[1 + \frac{\lambda\, C}
      {A(\bar{x}_{F_0})} \left( \frac{0.285  a }{\bar{x}^3_{F_0} }
+ \frac{1.350 b }{\bar{x}^4_{F_0}  } \right)  \right]\,.
\end{equation}

\begin{figure}[h!]
  \begin{center}
    \hspace*{-0.5cm} \includegraphics*[width=8.6cm]{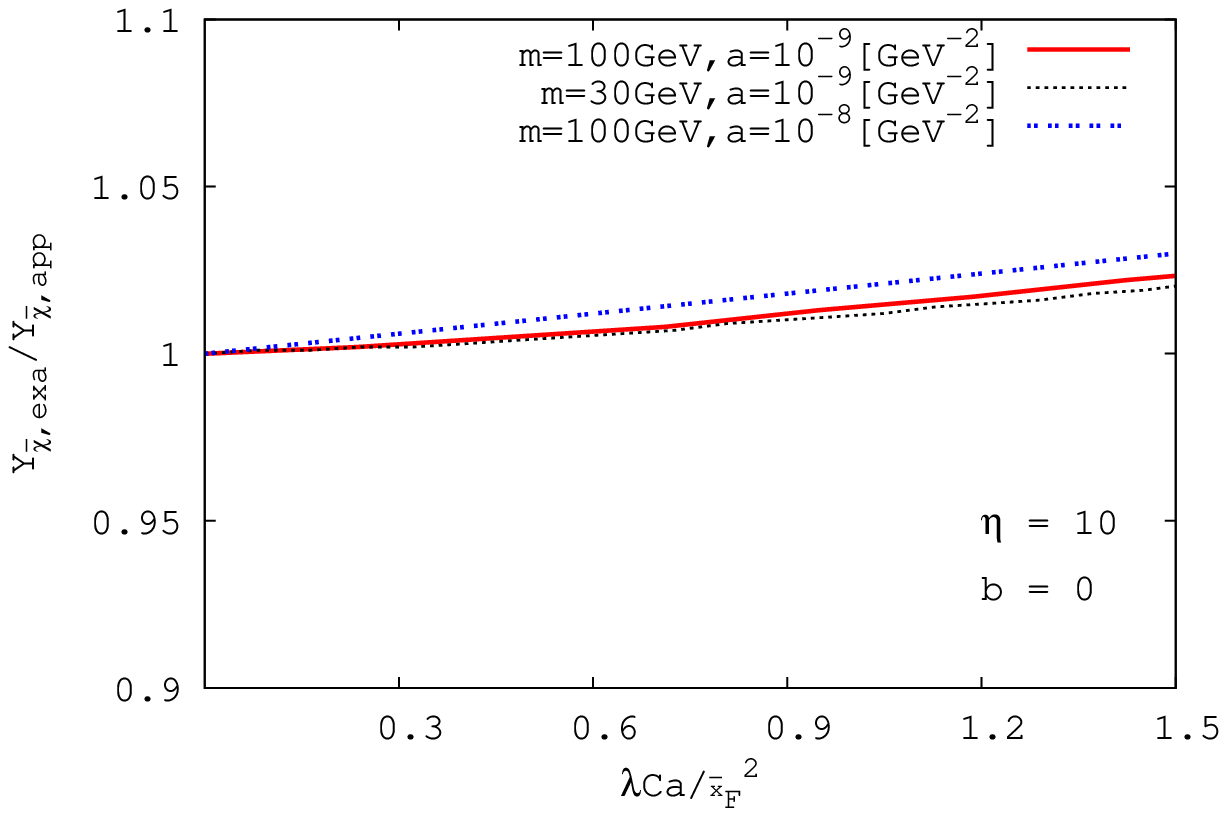}
    \hspace*{-0.5cm} \includegraphics*[width=8.6cm]{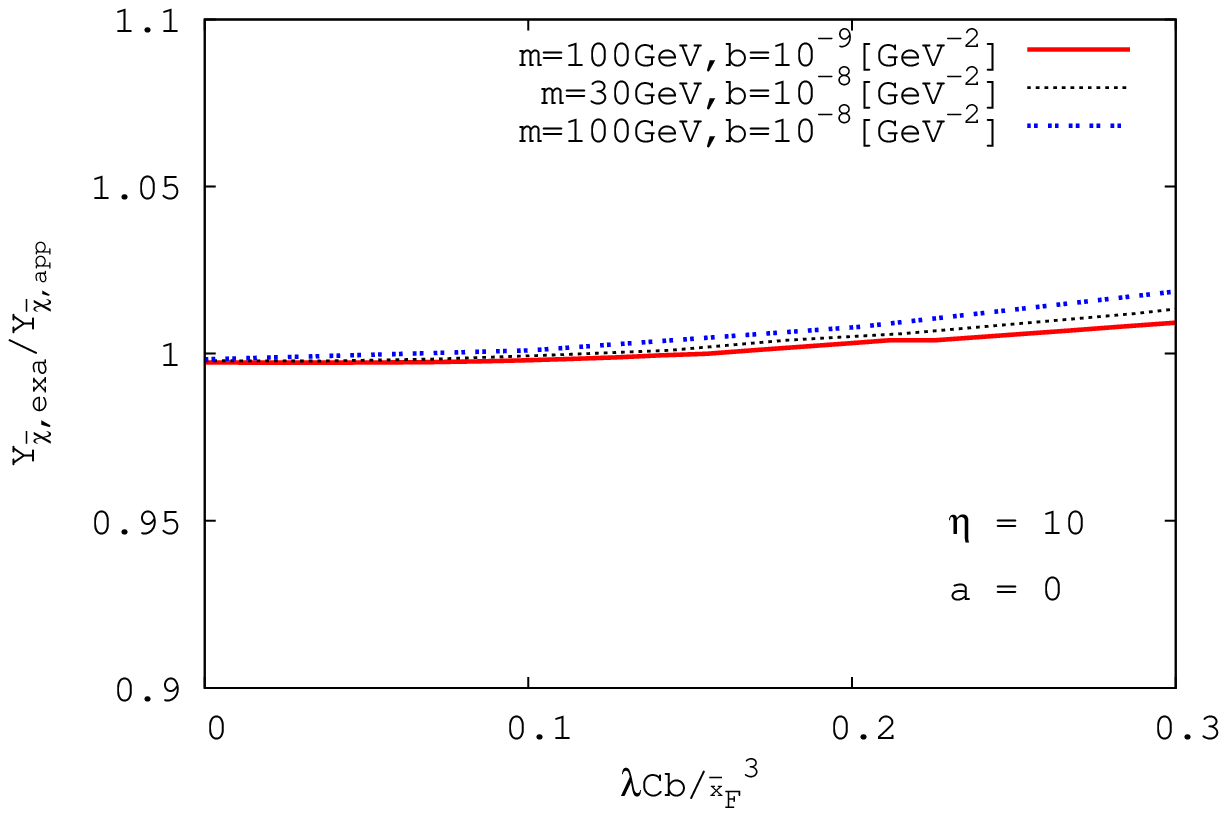}
    \caption{\label{fig:c} \footnotesize
    The ratio of the exact value of the $\bar\chi$
    particle abundance to the analytic value of $\bar\chi$
    particle abundance for $\eta = 10 $ (a) $b = 0 $ and (b) $a = 0$,
    $x_0 = 20$.  }
  \end{center}
\end{figure}
The ratio of the exact value of the $\bar\chi$ particle abundance to our
analytical approximation is plotted in Fig.\ref{fig:c} for quintessence
model with a kination phase. For
$\lambda C a /x_F^2 \lsim 1.5$ ( $\lambda C b /x_F^3 \lsim 0.3$ ),
the approximate analytic result matches the exact numerical result very well,
our approximation reproduces the exact numerical solution to better than
3\% (2\%) for $\eta = 10$.


\section{Constraints on Parameter Space}
\setcounter{footnote}{0}

The Dark Matter density is derived as in Eq.(\ref{wmap}) by WMAP team using
the CMB data for the minimal $\Lambda$CDM model. We use this result to find
the constraints on the enhancement factor $\eta$ in the quintessence model
with a kination phase. We use the following range for Dark Matter relic
density,
\begin{equation} \label{range}
0.10 < \Omega_{\rm DM} h^2 < 0.12
\end{equation}
The total Dark Matter density should be the addition of the particle $\chi$ and
anti--particle $\bar{\chi}$ contributions:
\begin{equation} \label{add}
\Omega_{\rm DM}=\Omega_\chi + \Omega_{\bar{\chi}}\,.
\end{equation}

\begin{figure}[t!]
  \begin{center}
    \hspace*{-0.5cm} \includegraphics*[width=8.7cm]{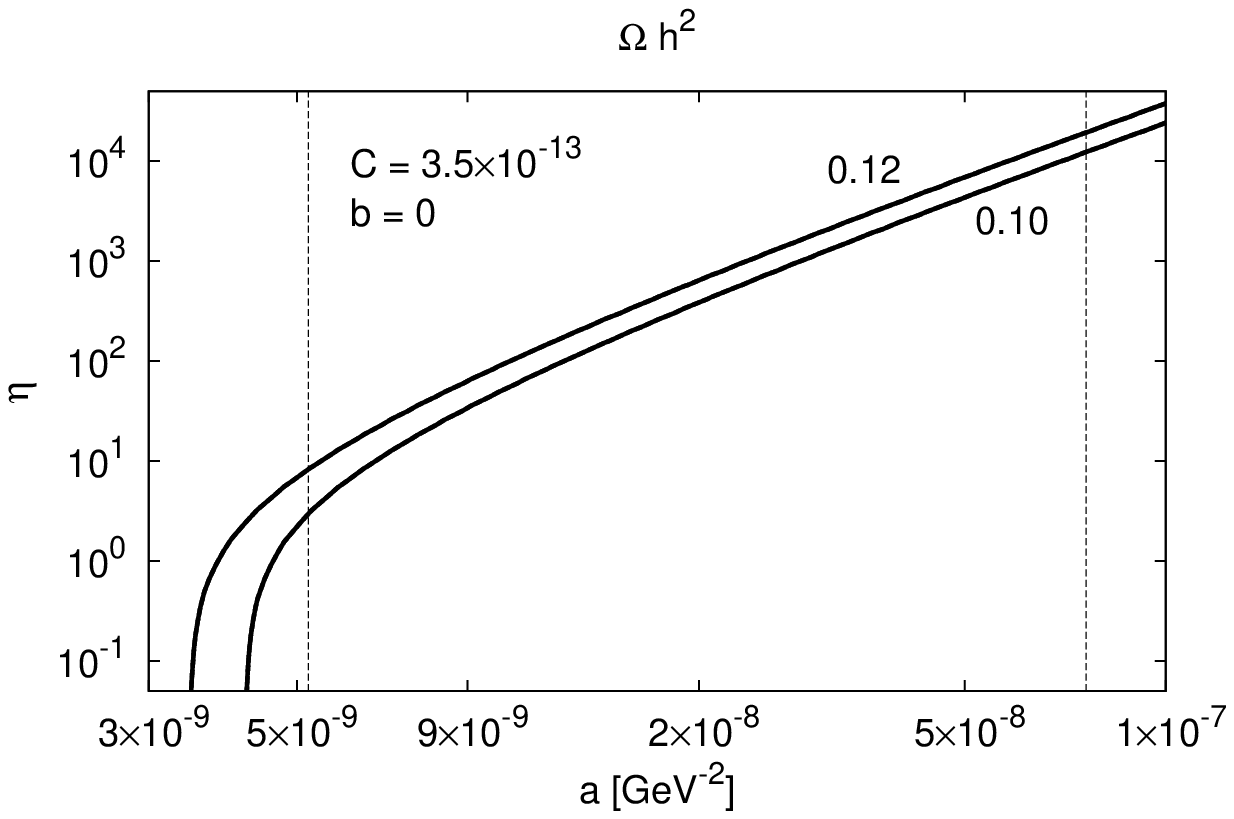}
    \put(-115,-12){(a)}
    \hspace*{-0.5cm} \includegraphics*[width=8.7cm]{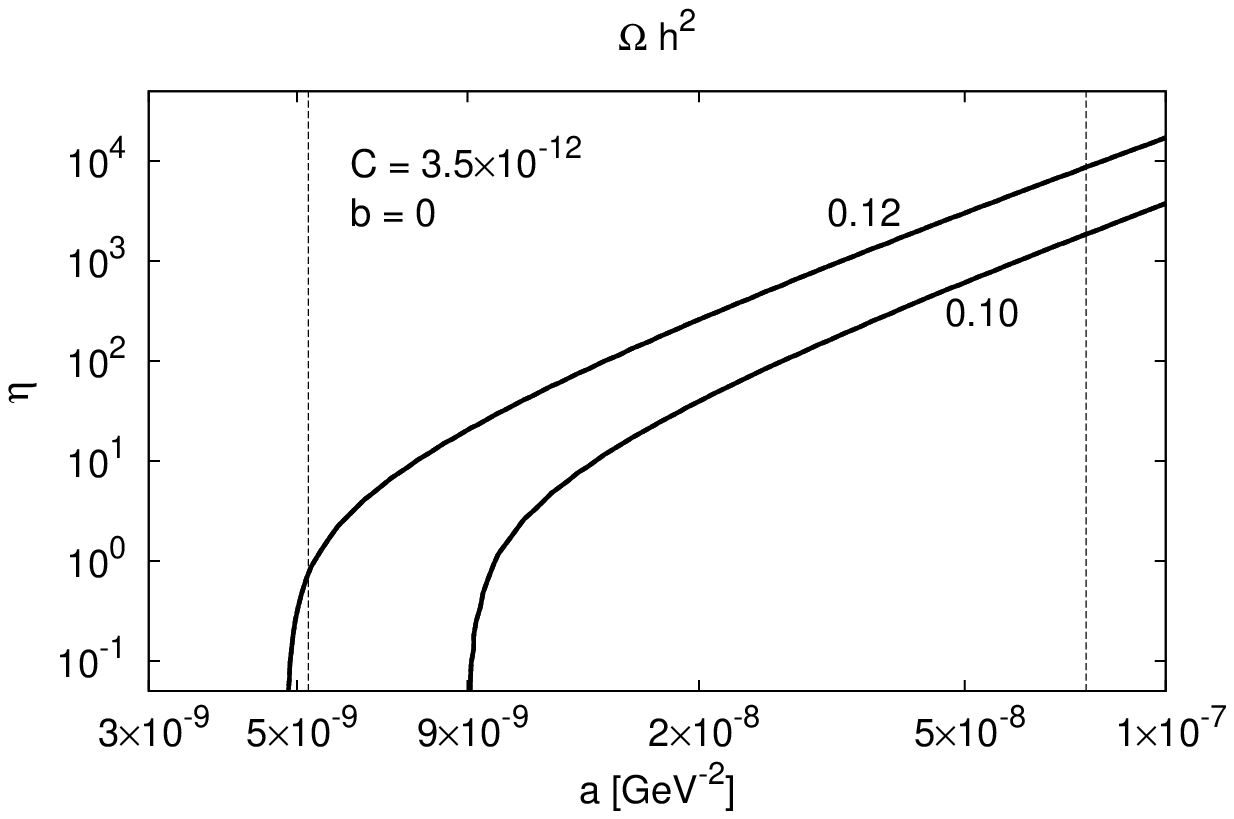}
    \put(-115,-12){(b)}
    \caption{\label{fig:d} \footnotesize
    The allowed region in the $(a,\eta)$ plane for $b=0$, when the
    Dark Matter density $\Omega h^2$ lies between 0.10 and 0.12. Here we
    take $m_\chi = 100$ GeV, $g_{\chi} = 2$ and $g_* = 90$, $x_0 = 20$; 
    $ C = 3.5 \times 10^{-13}$ for (a) and $C = 3.5 \times 10^{-12}$ for (b). The two vertical lines 
    $6.0 \times 10^{-26} $ cm$^3$ s$^{-1}$$=5.2 \times 10^{-9}$ GeV$^{-2}$ and
 $ 8.8 \times 10^{-25} $ cm$^3$ s$^{-1}$$=7.6 \times 10^{-8}$ GeV$^{-2}$ are the upper limits 
    for the cross sections for mass 100 GeV from the {\it Fermi}-LAT collaboration \cite{Ackermann:2011wa}. }
    \end{center}
\end{figure}
Fig.\ref{fig:d} shows the relation between the s--wave
annihilation cross section parameter $a$ and the enhancement factor
$\eta$ for two values of the total Dark Matter density. This figure is
based on the exact numerical solutions of Boltzmann equations
(\ref{eq:Ycq}), (\ref{eq:Ycbarq}).
We use the annihilation cross section which is given by
Eq.(\ref{eq:cross}). Here we take $m_\chi = 100$ GeV, $g_{\chi} = 2$
and $g_* = 90$, $x_0 = 20$; $ C = 3.5 \times 10^{-13}$ for (a) and $C =
3.5 \times 10^{-12}$ for (b). We choose such values for asymmetry factor $C$
which are in the range of the values to obtain the observed Dark Matter
abundance \cite{Iminniyaz:2011yp}. 

In \cite{Schelke:2006eg, Donato:2006af, Catena:2009tm}, the authors found that the maximal density 
enhancement compatible with BBN bounds is of the order of $10^6$ (for WIMP mass smaller than 1 TeV). 
In our work we used the observed Dark Matter abundance and derived constraints on
$\eta$ when the asymmetry factor $C$ is fixed. In the left frame of Fig.\ref{fig:d}, for small 
asymmetry factor $ C = 3.5 \times 10^{-13} $, it is shown
that for the values of s--wave annihilation cross sections from $a = 3.6 \times 10^{-9}$
GeV$^{-2}$ to $a = 1.0 \times 10^{-7}$ GeV$^{-2}$, the observed Dark Matter
abundance is obtained for the range of $\eta$ from $5.0 \times 10^{-2}$ to $3.8 \times10^4$.
In the right frame of Fig.\ref{fig:d}, for the asymmetry factor
$ C = 3.5 \times 10^{-12}$, one needs the s--wave annihilation cross sections
from $a = 4.8 \times 10^{-9}$ GeV$^{-2}$ to $a = 1.0 \times 10^{-7}$ GeV$^{-2}$ in the
 range of $\eta$ from $5.0 \times 10^{-2}$ to $1.5\times 10 ^4$ to obtain the observed Dark Matter
abundance. The annihilation cross section constraints are in the range of the
limit which are given by {\it Fermi}-LAT
collaboration \cite{Ackermann:2011wa}. {\it Fermi}-LAT
collaboration \cite{Ackermann:2011wa} gives the upper limit on the
cross section from about $10^{-26} $ cm$^3$ s$^{-1}$$= 8.6 \times 10^{-10}$ GeV$^{-2}$ at 5 GeV to
about $5 \times 10^{-23} $ cm$^3$ s$^{-1}$ $ = 4.3 \times 10^{-6}$ GeV$^{-2}$
at 1 TeV. For $m_{\chi} = 100$ GeV, the limit from {\it Fermi}-LAT collaboration \cite{Ackermann:2011wa}
is $6.0 \times 10^{-26} $ cm$^3$ s$^{-1}$$=5.2 \times 10^{-9}$ GeV$^{-2}$ to
$ 8.8 \times 10^{-25} $ cm$^3$ s$^{-1}$$=7.6 \times 10^{-8}$ GeV$^{-2}$.

\begin{figure}[t!]
  \begin{center}
    \hspace*{-0.5cm} \includegraphics*[width=8.7cm]{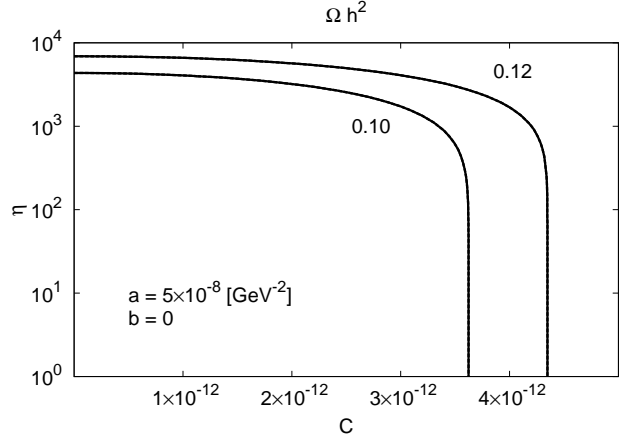}
    \caption{\label{fig:f} \footnotesize
    The allowed region in the $(\eta, C)$ plane for
    $a = 5 \times 10^{-8}$ GeV$^{-2}$, $b=0$, when the
    Dark Matter density $\Omega h^2$ lies between 0.10 and 0.12. Here we
    take $m_\chi = 100$ GeV, $g_{\chi} = 2$ and $g_* = 90$, $x_0 = 20$. }
    \end{center}
\end{figure}
Fig.\ref{fig:f} shows the allowed region in the $(\eta, C)$
plane for $a = 5 \times 10^{-8}$ GeV$^{-2}$, $b=0$, when the Dark Matter density
$\Omega_{\rm DM} h^2$ lies between 0.10 and 0.12.
Here we take $m_\chi = 100$ GeV,
$g_{\chi} = 2$ and $g_* = 90$, $x_0 = 20$. When the asymmetry factor $C$
ranges from $C = 0$ to $C = 3.6 \times 10^{-12}$, $\eta$ should be
around $\eta = 7.5 \times 10^2$ to $\eta = 8.0 \times 10^3$ to obtain
the observed Dark Matter relic density. $\eta$ is not sensitive to the
asymmetry factor in this case. In contrast for asymmetry factor which ranges
from $C = 3.6 \times 10^{-12}$ to $C = 4.3 \times 10^{-12}$, $\eta$ is from $\eta = 1$ to
$\eta = 7.5 \times 10^2$ to obtain the observed Dark Matter abundance.

\section{Summary and Conclusions}

In this paper we investigated the relic abundance of asymmetric WIMP
Dark Matter in quintessence model with a kination phase. The Dark
Matter particles and anti--particles are distinct in the asymmetric Dark
Matter Scenario. We assume the asymmetry starts well before the epoch of
thermal decoupling of the WIMPs. In quintessence model with a
kination phase, the asymmetric Dark Matter particles freeze--out earlier
than the standard case due to the enhanced Hubble rate. This leads to
the increase of the relic density of asymmetric Dark Matter particles. We
treat the enhancement factor $\eta$ and asymmetry factor $C$ as
free parameters in our work.

The discussion of the relic density of asymmetric Dark Matter
in the standard cosmological scenario which assumes the particles were in
thermal equilibrium in the early universe and decoupled when they were
non--relativistic has been done in paper \cite{Iminniyaz:2011yp, GSV}.
In our work, we extend it to the nonstandard cosmological scenario where the
Hubble rate of the universe is changed in quintessence model. We investigated
the relic density of asymmetric WIMP Dark Matter in this model. Using
the observed Dark Matter abundance, we find the constraints on the enhancement
factor $\eta$ assuming the asymmetry factor $C$ is fixed.

We found that the relic densities of both particles and
anti--particles are increased in quintessence model with a kination
phase. The size of the increase depends on the enhancement factor $\eta$.
For the large enhancement factor $\eta$, the increases
are more sizable than the smaller enhancement factor $\eta$.

For the range of cross section $a = 3.6 \times 10^{-9}$ GeV$^{-2}$ to
$a = 1.0 \times 10^{-7}$ GeV$^{-2}$, one needs the enhancement factor
$\eta$ from $5.0 \times 10^{-2}$ to
$3.8 \times 10^4$ for $C = 3.5  \times 10^{-13}$ to obtain the observed relic
density of asymmetric Dark Matter. When $C = 3.5 \times 10^{-12}$, the observed 
Dark Matter abundance is obtained for the enhancement factor $\eta$ from
 $ 5.0 \times 10^{-2}$ to $ 1.5 \times 10^4$ for the cross sections
from $a = 4.8 \times 10^{-9}$ GeV$^{-2}$ to $a = 1.0 \times 10^{-7}$ GeV$^{-2}$.

Our result is important to understand the relic abundance of
asymmetric Dark Matter in the early universe before Big Bang
Nucleosynthesis starts. From the observation we can also have
constraints on the parameter space for nonstandard cosmological
model in asymmetric Dark Matter case. In quintessence model with a
kination phase, the abundance of Dark matter and anti--Dark Matter
particles are increased. 

{\color{black}Note added}: as we were revising this manuscript, \cite{Gelmini:2013awa} was appeared, 
in which the asymmetric dark matter relic density is discussed in nonstandard cosmological 
scenarios including quintessence model with a kination phase and scalar--tensor model. 
While the general ideas are similar, the details of the treatment are different, 
in our work we only concentrated on the quintessence model with a kination phase.
For the kination model, our results are in general agreement with theirs.

\section*{Acknowledgments}

We thank professor Bi Xiao-jun and Shan-Chung Lin for useful discussions.
The work is supported by the National Natural Science Foundation of China
(11047009) and by the doctor fund BS100108 of Xinjiang university.

\end{document}